\documentclass{article}

\usepackage{amsmath}
\usepackage{feynmf}

\input epsf

\begin{document}

\setlength{\unitlength}{1mm}

\begin{titlepage}

\begin{flushright}
LAPTH-820/00\\
Edinburgh 2000/28\\
December 2000
\end{flushright}
\vspace{1.cm}

\begin{center}
\large\bf
{\LARGE \bf Beyond leading order effects in photon pair 
production at the Tevatron}\\[2cm]
\rm
{ T.~Binoth$^{a,b}$, J.~Ph.~Guillet$^{a}$, E.~Pilon$^{a}$ and M. Werlen$^{a}$}
\\[.5cm]

{\em $^{a}$ Laboratoire d'Annecy-Le-Vieux de Physique 
Th\'eorique\footnote{UMR 5108 du CNRS, associ\'ee \`a l'Universit\'e de Savoie.}
LAPTH,}\\
{\em Chemin de Bellevue, B.P. 110, F-74941  Annecy-le-Vieux, France}\\[.2cm]
{\em $^{b}$ Department of Physics and Astronomy,
University of Edinburgh, }\\
{\em  Edinburgh EH9 3JZ, Scotland}        
      
\end{center}
\normalsize

\vspace{2cm}

\begin{abstract}
We discuss effects induced by beyond leading order contributions to photon pair
production. We point out that next to leading order contributions to the
fragmentation component of the signal lead to a change in the shape of
distributions. This is already mildly visible in present Tevatron data though
stringent isolation criteria tend to suppress the fragmentation component
considerably. We expect the effect to be experimentally confirmed in future
data samples with higher statistics which would serve as a precision test for
QCD. 
\end{abstract}

\vspace{3cm}

\end{titlepage}

\section{Introduction}
The production of prompt photon pairs with large invariant mass is the 
object of continuing experimental studies ranging from fixed target 
\cite{wa70cross,wa70correl,e706}, to colliders energies \cite{ua2,cdf,d0}, 
especially by the CDF and D0 experiments at the Tevatron. The word ``prompt" 
means that these photons do not come from the decays of hadrons such as
$\pi^{0}$, $\eta$, etc. at large transverse momentum. Prompt photons may be
produced according to two possible mechanisms: either they take part directly
to the hard subprocess (direct mechanism), or they result from the
fragmentation of large transverse momentum partons (fragmentation mechanism).
For a general discussion of the production mechanisms of prompt photon pairs, 
see Ref. \cite{bgpw}. Besides offering an interesting probe of the
short distance QCD dynamics, di-photon hadroproduction is the irreducible
background for the search of a neutral Higgs boson in the channel 
$H \rightarrow \gamma \gamma$ in the mass range 90~-~140~GeV at the LHC, which
provides a strong motivation for its extensive study.

\vspace{0.2cm}

Collider experiments at the Tevatron and the forthcoming LHC do not perform
{\it inclusive} photon measurements. Indeed, the inclusive production rates of
$\gamma \pi^{0}$ and $\pi^{0} \pi^{0}$ pairs with large invariant mass 
are orders of 
magnitude 
larger than the one of prompt photon pairs, to which they provide 
the background. Thus the experimental selection of prompt photons
requires the use of isolation cuts. The isolation criterion used by the
Tevatron experiments CDF \cite{cdf} and D0 \cite{d0} is schematically the
following. A photon is said to be isolated if, in
a cone in rapidity and azimuthal angle about the photon direction, the amount 
of deposited hadronic transverse energy $E_{T}^{had}$ is smaller than some 
value $E_{T \, max}$ fixed by the experiment:
\begin{equation}\label{isol}
E_{T}^{had} \leq E_{T \, max} \;\;\;\; \mbox{inside} \;\;\;\; 
\left(  y - y_{\gamma} \right)^{2} +  
\left(  \phi - \phi_{\gamma} \right)^{2} \leq R^{2} 
\end{equation}
The actual experimental criteria, which are defined at the detector level, are
much more complicated than the simple one given by Eqn. (\ref{isol}), see
\cite{cdf,d0}. They cannot be implemented in a partonic calculation. The
schematical criterion of Eqn. (\ref{isol}) is a modelization of their effects
at the parton level. The topic of isolation of photons based on the above
criterion is extensively discussed in the literature 
\cite{boo1,abf,berger-qiu,gordon-vogelsang,vogt-vogelsang,cfgp}. 
Besides 
the rejection of the
background of secondary photons, the isolation cut also reduces the prompt 
photons from fragmentation. The stringent isolation requirements used by CDF
and D0 suppress the fragmentation component quite a lot; therefore, the effects
of the fragmentation component are generally neglected beyond their lowest order
in perturbation theory. Yet they can affect various observables in a
sizeable way. For example, in the case of single photon  production, it yields
a significant increase in the transverse momentum distribution in the lower
range of the $p_{T}$ spectrum \cite{gordon-vogelsang,vogt-vogelsang,cfgp}. 

\vspace{0.2cm}

In a recent work, we presented a full next to leading order (NLO) study of  
di-photon hadroproduction based on a computer code of partonic event generator
type, {\em DIPHOX} \cite{bgpw}. In this short article, relying on this general 
study, we illustrate the impact of beyond leading order contributions,
especially to the fragmentation component, in di-photon production at the
Tevatron.

\section{Impact of higher order corrections on di-photon spectra}

Higher order QCD contributions to a given observable lead to a refinement of
theoretical predictions in several aspects. If already the Born level contains
the strong coupling $\alpha_s$, higher orders stabilize dependencies on
arbitrary scales stemming from the truncation of the perturbative series. In
the hadronic production of direct photon pairs this is not  the case though it
is in general true for the  fragmentation parts. Furthermore, the inclusion of
the  NLO contributions allow for new initial states. The additional production
channels are typically the reason  for large $K$ factors. But higher order
contributions also modify final state observables, as extra particles in the
final state generate kinematical configurations which are forbidden at leading
order and lead to sizeable effects\footnote{Higher order corrections can also
reveal new infrared sensitive issues, in particular {\it inside} the physical 
spectrum, as it is the case for the $q_{T}$ distribution at 
$q_{T} = E_{T \; max}$ \cite{bgpw}.}, especially if new collinear situations in
the final states are possible.
Such effects can be seen in observables sensitive to kinematical 
configurations where the photons are emitted in rather close directions to 
each other. Indeed, such configurations contribute only beyond leading order
(at lowest order, transverse momentum conservation forces the photons to be
back to back in the transverse plane). Moreover, the contribution of such
configurations coming from the fragmentation mechanism is enhanced, as will be
discussed below.

\vspace{0.2cm}

\begin{figure}[p]
\hspace{2cm}
    \epsfxsize = 13cm
    \epsffile{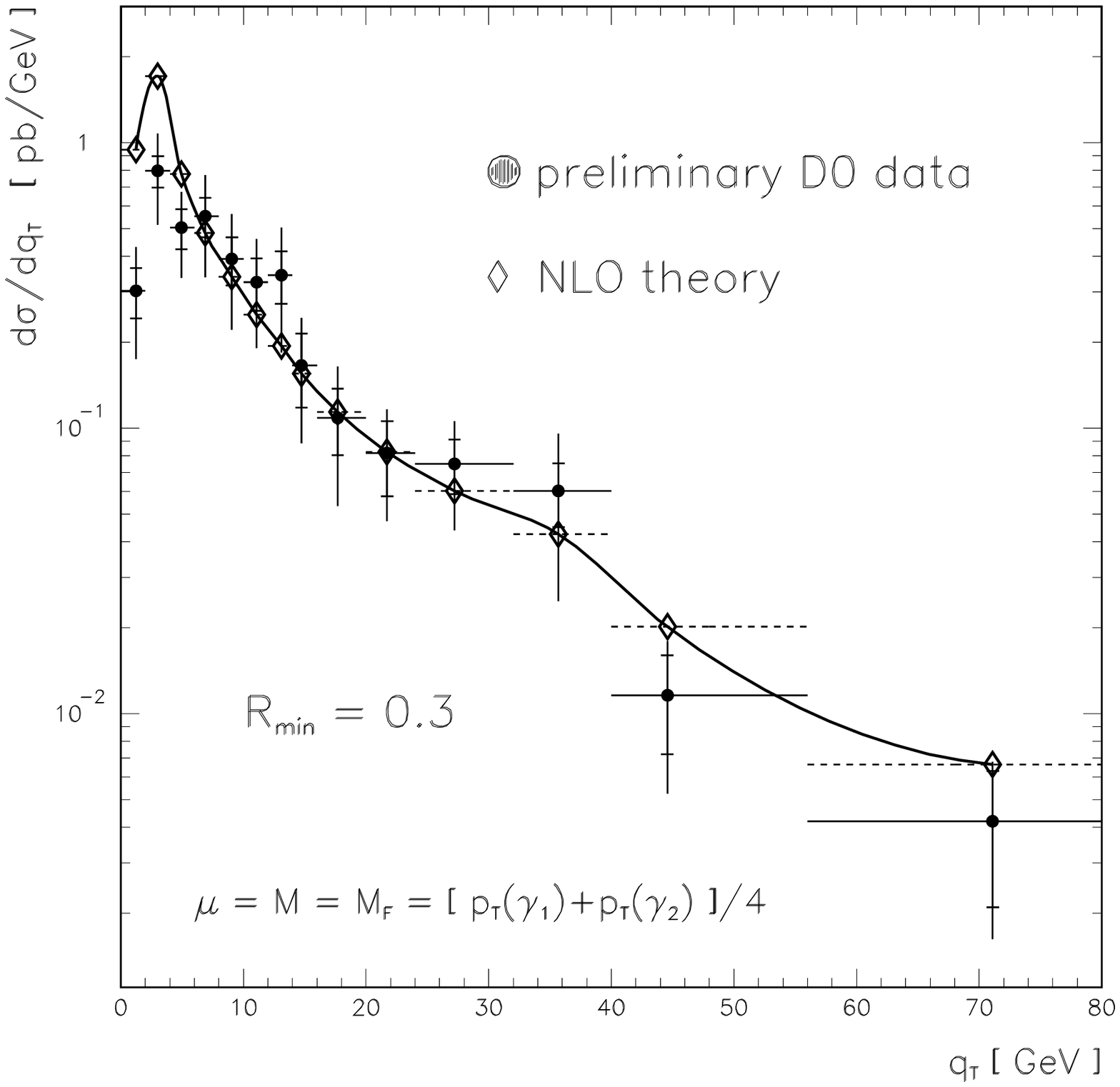}
\caption{\label{Fig:d0_qt}{\em $q_{T}$ distribution of photon pairs. 
Black dots: D0 data \cite{d0}; white diamonds: average values of the NLO 
calculation (DIPHOX code) in the corresponding experimental bins.
The curve is a spline interpolation between the theoretical average.}}
\end{figure}

A typical representative of this familly of observables is the distribution 
$d \sigma/ d q_{T}$ of transverse momentum of di-photons ($q_{T} = ||{\vec{p}}_{T}(\gamma_1) + {\vec{p}}_{T}(\gamma_2)||$). 
Fig. \ref{Fig:d0_qt} shows the comparision between the D0 preliminary data and 
our
theoretical computation taking into account the kinematic and isolation cuts of
the D0 Tevatron experiment. D0 requires $|y(\gamma_{1,2})| < 1.0$ for the photon rapidities, and the lower
experimental $p_{T}$ cuts are
\begin{equation}\label{p-t-cuts1}
p_{T}(\gamma_{1}) \geq 14 \; \mbox{GeV} ,\;\;\;\; p_{T}(\gamma_{2}) \geq 13 \; \mbox{GeV}
\end{equation} 
which roughly correspond to effective cuts at respectively
\begin{equation}\label{p-t-cuts2} 
p_{T \, min}(\gamma_{1}) = 14.90 \; \mbox{GeV}, \;\;\;\; 
p_{T \, min}(\gamma_{2}) = 13.85 \; \mbox{GeV} 
\end{equation} 
in a partonic calculation\footnote{The experimental cuts at measured values 14 
and 13 GeV respectivelly in the D0 data are not corrected for the 
electromagnetic calorimeter absolute energy scale. See \cite{bgpw} for more 
details.}. Furthermore, a lower cut in the two-photon acollinearity is required:
\begin{equation}\label{acolln} 
\left( y(\gamma_{1}) - y(\gamma_{2}) \right) ^{2} + \phi_{\gamma \gamma}^{2} 
\geq R_{min}^{2}
\end{equation}
where $\phi_{\gamma \gamma}$ is the photon-photon azimuthal angle and 
$R_{min} = 0.3$. The isolation criterion is modelized according to  
Eqn. (\ref{isol}) with $E_{T \, max} = 2$ GeV, $R = 0.4$. 

\vspace{0.2cm}

One can see a shoulder at about $q_{T} \sim$ 35 GeV in the distribution measured
experimentally. At first glance, one may question its significance, given the
limited statistics of the Run I data and the systematic uncertainties which
affect these preliminary data. However a corresponding shoulder appears in the
theoretical calculation performed with the same binning as the data too. A
carefull study reveals that this shoulder is not a mere statistical
fluctuation of the partonic Monte-Carlo generator, nor a binning artefact, but
it is a physical effect. 

\vspace{0.2cm}

In order to see this, we split the two-photon phase space into regions, 
according to the photon-photon azimuthal angle $\phi_{\gamma \gamma}$: 
\begin{equation}\label{slicing}
\mbox{Region  I:}  \;\;\;\; 0 \leq \phi_{\gamma \gamma} < \frac{\pi}{2}; 
\;\;\;\;
\mbox{Region II:}  \;\;\;\;\frac{\pi}{2} \leq \phi_{\gamma \gamma} \leq \pi.
\end{equation}
Since at leading order, the two body kinematics forces the photons to be 
back to back in the transverse momentum plane ($\phi_{\gamma \gamma} = \pi$), 
the leading order contributions to $d \sigma/ d q_{T}$ come entirely from 
region II, and vanish in region I. 
With respect to the above artificial slicing (\ref{slicing}), a ``new channel" 
opens at NLO - and beyond - which populates region I. However, as 
$q_{T} = ||{\vec{p}}_{T}(\gamma_1) + {\vec{p}}_{T}(\gamma_2)||$, 
the asymmetric cuts on the $p_{T}$'s of the photons imply that the contribution 
to $d \sigma/ d q_{T}$ from region I vanishes for $q_{T} \leq q_{T \, min}$ 
with $q_{T \, min} = (p_{T \, min }^{2}(\gamma_1) + p_{T \, min}^{2}(\gamma_2))^{1/2}$, so 
that this ``new channel" opens only above $q_{T \, min} \simeq 20.34$ GeV. It 
is instructive to examine separately the direct, and the fragmentation 
components\footnote{We call {\it direct} the mechanism by which both photons
take part to the hard subprocess, {\it one fragmentation} the one by which one
photon takes part to the hard subprocess whereas the other comes from 
fragmentation, and {\it two fragmentation} the one by which both photons come
from fragmentation. We call {\it fragmentation} the sum of these last two
components. Actually, the severe isolation cut used by D0 suppresses  the 
{\it two fragmentation} component down to a completely negligible level. 
Consequently, we will ignore it in the present discussion.}\label{ftnt1} above 
this threshold. 

\vspace{0.2cm}

\begin{figure}[p]
\hspace{2cm}
    \epsfxsize = 13cm
    \epsffile{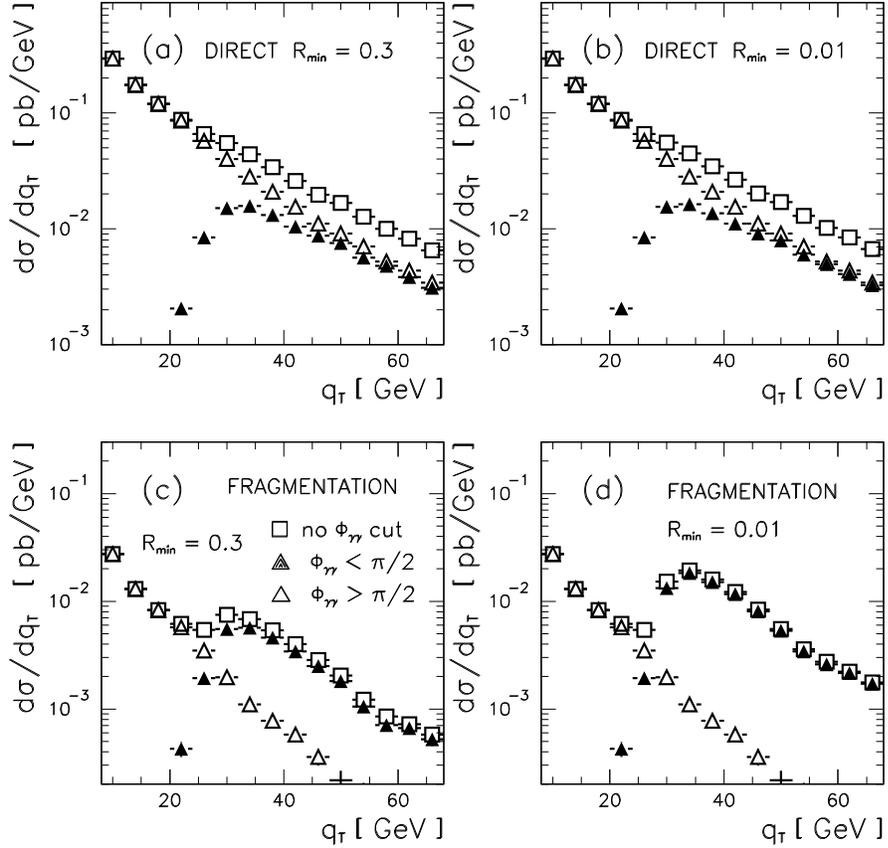}
\caption{\label{Fig:qt_shoulder}{\em Origin of the 
$q_{T}$ shoulder in the theoretical calculation. Plot (a) shows the direct component (open squares) split into the phase space regions $\phi_{\gamma \gamma}< \pi/2$ (full triangles) and $\phi_{\gamma \gamma}> \pi/2$ (open triangles) for the experimentally used value $R_{min} = 0.3$. Plot (b) is the same as (a) but for $R_{min} = 0.01$ to show the sensitivity of the effect to this collinear cut. Plots (c) and (d) show the corresponding histograms for the fragmentation component where the enhancement due to photon collinearity is clearly visible.}}
\end{figure}

The direct component is displayed on Fig. \ref{Fig:qt_shoulder}(a). As 
$q_{T}$ increases above $q_{T \, min}$, the size of the
opening slice of  available phase space belonging to region I becomes
comparable with the one of region II. The direct component, which starts at
NLO for  $q_{T} \geq q_{T \, min}$, is by definition free of any final state
collinear singularity. It thus involves only smooth variations with  
$\left( y(\gamma_{1}) - y(\gamma_{2}) \right)$ and $\phi_{\gamma \gamma}$. 
Therefore, after a transition  
regime extending up to $q_{T} \leq 35-40$~GeV, the magnitudes of the
contributions from regions I and II become comparable. Moreover the
$R_{min}$ dependence, which comes from region I, is weak, 
${\cal O}(R_{min}^{2})$ at small $R_{min}$. In summary the approximate doubling 
of available phase space due to the opening of region I above $q_{T \, min}$
roughly amounts to a doubling of the direct component.

\vspace{0.2cm}

\begin{figure}[p]
\begin{center}
\parbox[c][80mm][c]{80mm}{\begin{fmfgraph*}(80,80)
  \fmfleftn{i}{1} \fmfrightn{o}{1}
  \fmftopn{t}{9} \fmfbottomn{b}{9}
  \fmfcmd{%
    vardef middir(expr p,ang) =
      dir(angle direction length(p)/2 of p + ang)
    enddef;
    style_def arrow_left expr p =
      shrink(.7);
        cfill(arrow p
	shifted(4thick*middir(p,90)));
      endshrink
    enddef;
    style_def arrow_right expr p =
      shrink(.7);
        cfill(arrow p
	shifted(4thick*middir(p,-90)));
      endshrink
    enddef;}
  \fmfcmd{%
    vardef middir(expr p,ang) =
      dir(angle direction length(p)/2 of p + ang)
    enddef;
    style_def warrow_left expr p =
      shrink(.7);
        cfill(arrow p
	shifted(8thick*middir(p,90)));
      endshrink
    enddef;
    style_def warrow_right expr p =
      shrink(.7);
        cfill(arrow p
	shifted(8thick*middir(p,-90)));
      endshrink
    enddef;}
  \fmf{plain,tension=2.}{i1,v1}
  \fmf{arrow_right}{i1,v1}
  \fmf{gluon,tension=1.}{o1,v1}
  \fmf{warrow_left}{o1,v1}
  \fmf{photon,tension=2.,label.side=right,label=$\gamma_{1}$}{t6,v3}
  \fmf{arrow_right}{v3,t6}
  \fmf{plain,tension=2.}{v1,v2}
  \fmf{plain,tension=1.4}{v2,v3}
  \fmf{gluon,tension=1.43}{v1,fv1}
  \fmf{warrow_left}{v1,fv1}
  \fmf{phantom,tension=1.5}{fv1,b7}
  \fmfv{decor.shape=circle,decor.filled=shaded,decor.size=.05w,
  label=$D_{\gamma/q}$,label.angle=180.,label.dist=0.05w}{v3}
  \fmfv{decor.shape=circle,decor.filled=shaded,decor.size=.08w,
  label=$\sigma_{q g \rightarrow q g}$,label.angle=150.,label.dist=0.05w}{v1}
  \fmf{photon,tension=1.}{v2,fv2}
  \fmf{arrow_right}{v2,fv2}
  \fmfv{label=$\gamma_{2}$,label.angle=-50.,label.dist=0.03w}{fv2}
  \fmf{phantom,tension=1.}{fv2,t8}
\end{fmfgraph*}}
\end{center}
\caption{\label{Fig:kinematics}{\em Kinematical configuration for which the
fragmentation contribution is collinearly enhanced when the two photons become close to each other.}}
\end{figure}
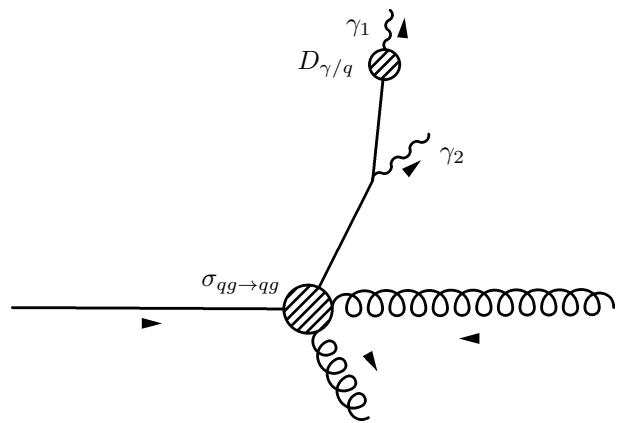

On the contrary, the fragmentation component is collinearly enhanced when the
two photons come close to each other. Indeed, without the cut (\ref{acolln}), 
contributions such as the one pictured in Fig. \ref{Fig:kinematics} would make 
the fragmentation component logarithmically divergent for all $q_{T}$ larger 
than $q_{T\; lim}= p_{T \, min}(\gamma_{1}) + p_{T \, min}(\gamma_{2})$ with 
$q_{T\; lim} \simeq 28.75$~GeV. This divergence would be produced when the 
photon considered as ``direct" would become collinear to the one from  
fragmentation\footnote{From a theoretical point of view, the subtraction of 
the corresponding double collinear divergence for the \{partonic emitter + two
photons\} system in the partonic calculation would require a new type of
fragmentation functions of partons into ``twin collinear photons", extending
the procedure corresponding to the case of fragmentation into a single photon. 
Such collinear photons would actually be detected as single photon events. 
Most probaby they would contribute to the single photon production rate at a 
negligeable level, at least in the Tevatron energy range, due to a relative suppression factor $\propto {\cal O}(\alpha_{em})$ with respect to the typical single photon production rate. In practice, the experimental cuts on the acollinearity $(y(\gamma_{1})-y(\gamma_{2}))^{2} + \phi_{\gamma \gamma}^{2}$ 
of the two photons, or on the 
invariant mass of the pair, prevent from having to consider this regime.}. 
The acollinearity cut (\ref{acolln}) forbids this configuration; yet a
collinear enhancement survives in region I as a memory of this regime. As the
slice of available phase space in region I opens, the fragmentation
contribution from this region rises initially faster than the contribution from
region II decreases above $q_{T \, min}$. This generates a peak in the
fragmentation component about $q_{T} \sim$ 30-35 GeV, untill the contribution
from region I decreases too. As a reflect of the collinear enhancement, the
smaller the value of $R_{min}$, the higher the peak. For $q_{T \, min}$ above 
25-30 GeV, the fragmentation term comes mostly from region I, contrarily
to the direct case. This is illustrated on Fig. \ref{Fig:qt_shoulder} which 
compares the case $R_{min} = 0.3$ experimentally used and what would happen for 
$R_{min} = 0.01$, an artifical value chosen on purpose to emphasize this effect. 

\vspace{0.2cm}

\begin{figure}[p]
\hspace{2cm}
    \epsfxsize = 13cm
    \epsffile{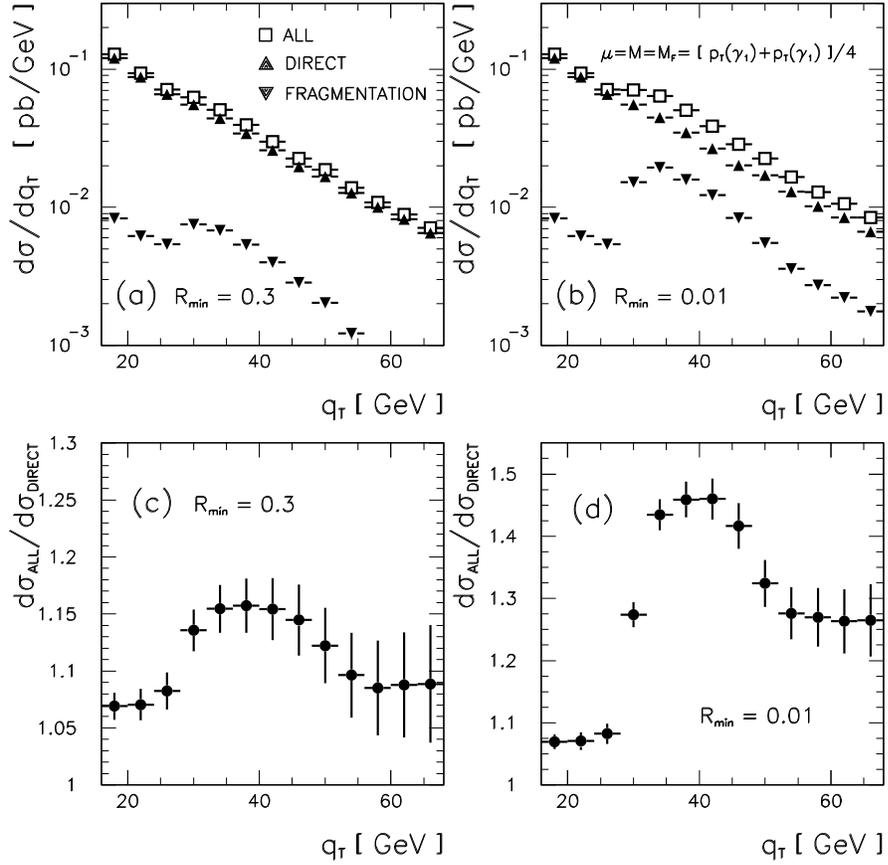}
\caption{\label{Fig:qt_rggcut_eb}{\em Top: the $q_{T}$ spectrum (open squares) split into direct (full triangles) and fragmentation part (inversed triangles) for $R_{min} = 0.3$ (a) and $R_{min} = 0.01$ (b). Bottom: the ratio of the total $q_{T}$ distribution divided by the direct one for $R_{min} = 0.3$ (c) and $R_{min} = 0.01$ (d).}}
\end{figure}

Due to the fact that the fragmentation component is severely suppressed by the 
isolation cut~(\ref{isol}) with respect to the direct one, and that the value 
$R_{min} = 0.3$ experimentally used leads to a rather moderate enhancement
of this fragmentation component, the effect of the latter is limited to a  
10\% modulation effect in the total (= direct + fragmentation) distribution,
so that the shoulder is mainly due to the direct component. This can be seen
in Fig. \ref{Fig:qt_rggcut_eb} showing the ratio total/direct. However,
the effect would be more spectacular with lower values of $R_{min}$. For
illustrative purpose, Fig. \ref{Fig:qt_rggcut_eb} also shows what would be the 
situation with the value $R_{min} = 0.01$.

\vspace{0.2cm}

\begin{figure}[p]
\hspace{2.cm}
    \epsfxsize = 13cm
    \epsffile{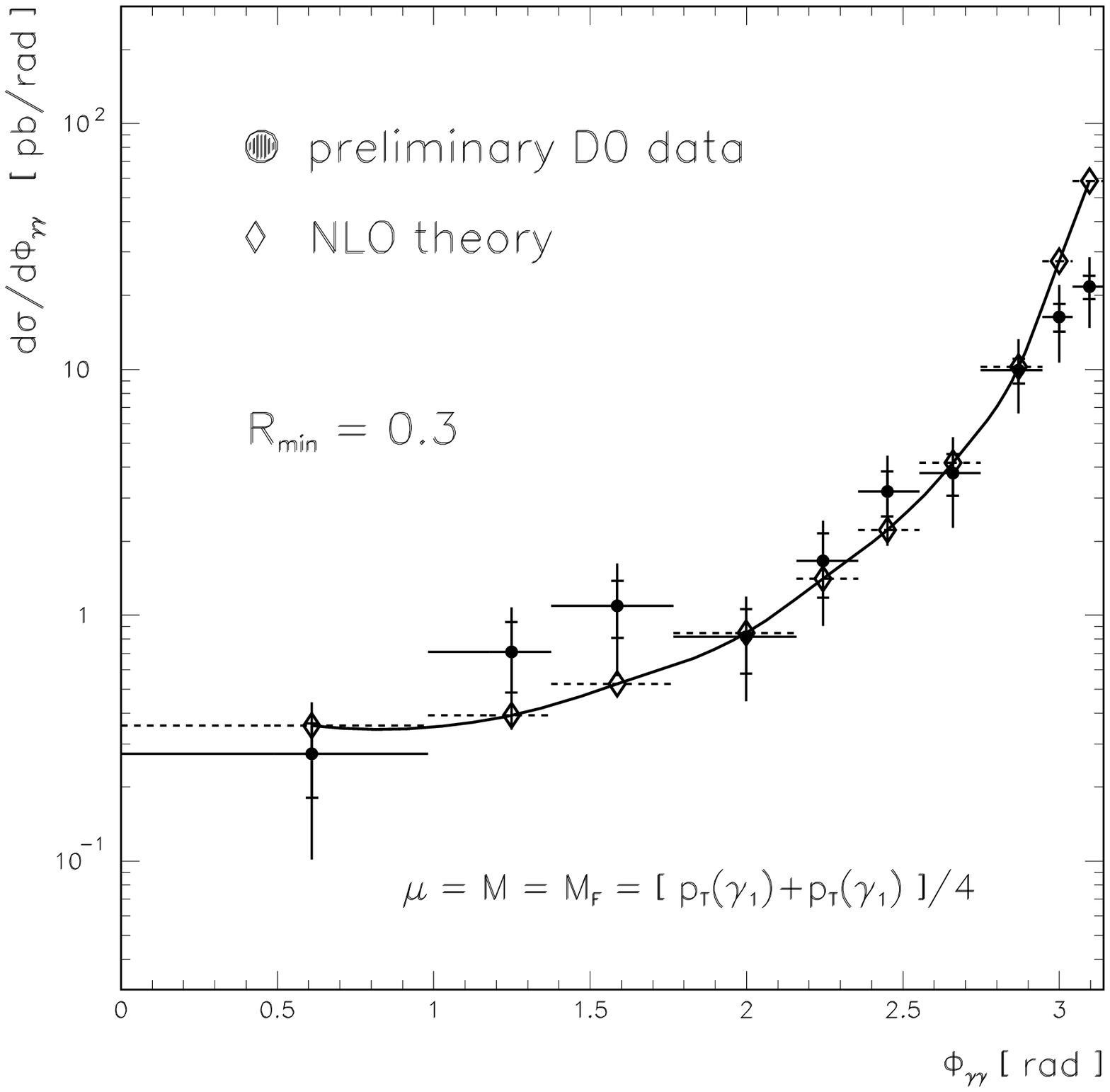}
\caption{\label{Fig:d0_phi}{\em $\phi_{\gamma \gamma}$ distribution of photon
pairs. Black dots: D0 data \cite{d0}, white diamonds: average values of the NLO 
calculation (DIPHOX code) in the corresponding experimental bins.
The curve is a spline interpolation between the theoretical average.}}
\end{figure}

\begin{figure}[p]
\hspace{2cm}
    \epsfxsize = 13cm
    \epsffile{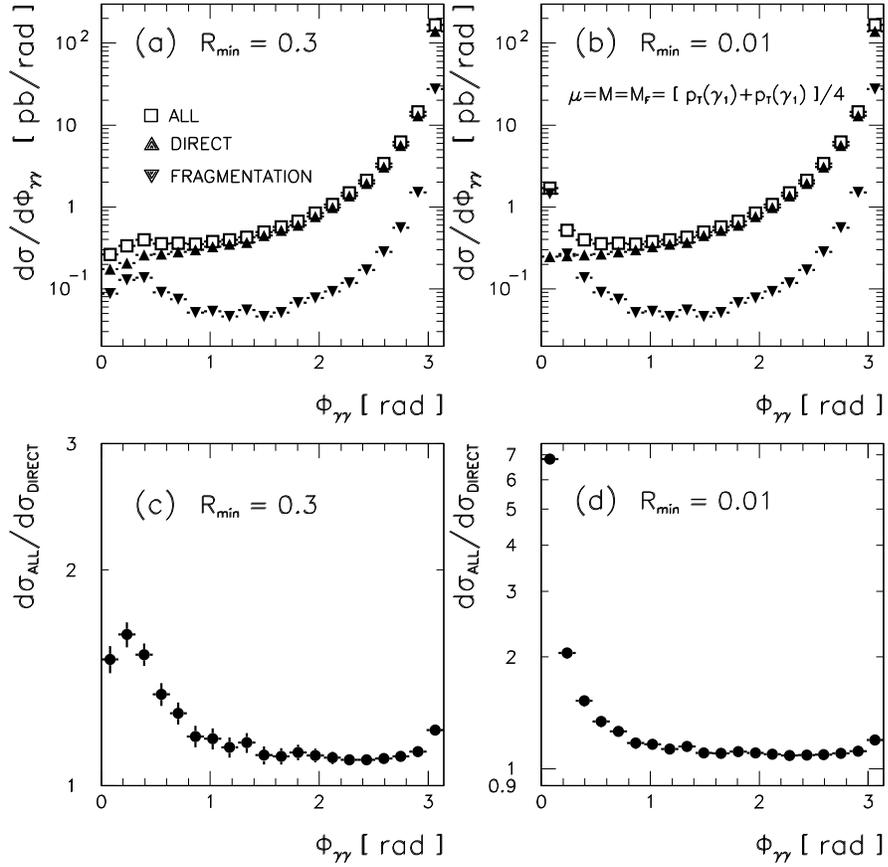}
\caption{\label{Fig:phi_rggcut_eb}{\em Top: the theoretical $\phi_{\gamma \gamma}$ distribution (open squares) split into direct (full triangles) and fragmentation part (inversed triangles) for $R_{min} = 0.3$ (a) and $R_{min} = 0.01$ (b). Bottom: the ratio of the total $\phi_{\gamma \gamma}$ distribution divided by the direct one for $R_{min} = 0.3$ (c) and $R_{min} = 0.01$ (d). The collinear enhancement of the fragmentation part for small $\phi_{\gamma \gamma}$ is clearly visible.}}
\end{figure}

The collinear enhancement in the fragmentation component affects also the
distribution $d \sigma/d \phi_{\gamma \gamma}$ of azimuthal angle between the 
two photons in a pair, in the lower range of the $\phi_{\gamma \gamma}$ 
spectrum, cf. Fig. \ref{Fig:d0_phi}. 
The contribution of the direct component decreases 
monotonically while remaining finite when $\phi_{\gamma \gamma} < \pi$ 
decreases towards 0. Instead, the fragmentation contribution to this observable 
would diverge logarithmically at $\phi_{\gamma \gamma} =$ 0 without the 
acollinearity cut (\ref{acolln}), as can be inferred from Fig. \ref{Fig:phi_rggcut_eb}. The fragmentation contribution 
``feels" this collinear divergence by rising again when $\phi_{\gamma \gamma}$ 
decreases below $\sim 1$ towards $R_{min}$. This divergence is forbidden by the 
constraint (\ref{acolln}). The latter produces a turnover when 
$\phi_{\gamma \gamma} \leq R_{min}$, which can be seen on the right plot on the 
right of Fig. \ref{Fig:phi_rggcut_eb} in the lower range of the spectrum.

\vspace{0.2cm}
One may wonder about the significance of these effects compared with the
theoretical scale uncertainties associated with the arbitrariness of the
renormalization and factorization scales used in the partonic
calculation\footnote{For definiteness, we used the common scale 
$M = M_{F} = \mu = (p_{T}(\gamma_{1}) +  p_{T}(\gamma_{2}))/4$ for the initial
state factorization, final state fragmentation, and renormalization scales.}.
Concerning the $q_{T}$ distribution, the direct contribution to the tail of 
the distribution above $q_{T \; min}$ starts at NLO (the Born contribution 
to the direct component is concentrated at $q_{T} =0$ by transverse momentum 
conservation). As for the $\phi_{\gamma \gamma}$ distribution, both the direct 
and fragmentation contributions to the tail of the distribution away from 
$\phi_{\gamma \gamma} = \pi$ start at NLO too (the respective Born terms being 
concentrated at $\phi_{\gamma \gamma} = \pi$). For each of these two 
observables, NLO is the {\it effective} lowest order. Consequently, it is 
fair to say that these uncertainties are rather large. 
However scale uncertainties act roughly as an overall effect: they do not
generate local distortions comparable to the pattern of effects studied above.
We insist that the latter are {\it physical} effects, which could in principle
be  unambiguously observed among data with higher statistics and an improved 
understanding of systematics, as it may hopefully be the case in the 
forthcoming Run II of the Tevatron.

\section{Conclusions and perspectives.}
In this note we have emphasized NLO effects, coming in particular from the
fragmentation contribution to di-photon production. We have pointed out that a
new collinear sensitive situation arises in the fragmentation part at
next to leading order, namely when the two photons become more and more close
to each other. This leads to an enhancement of the respective signal. 
Stringent isolation criteria reduce the fragmentation contribution considerably
and acollinearity cuts between the two photons forbid collinear photons in the
experimental situation. Consequently, these next to leading order effects from
fragmentation are not very significant in the present data. However we have
shown that the account of these contributions is required for an accurate
description of the shapes of di-photon observables. They thus may become
relevant for a precise understanding of the high statistic data to be collected
during the forthcoming Run II at the Tevatron.

\vspace{1.0cm}
\noindent
{\it \bf Acknowledgments.} This work was supported in part by the EU Fourth
Training Programme ``Training and Mobility of Researchers", Network ``Quantum
Chromodynamics and the Deep Structure of Elementary Particles", contract
FMRX-CT98-0194 (DG 12 - MIHT). LAPTH is a Unit\'e Mixte de Recherche 
(UMR 5108) du CNRS associ\'ee \`a l'Universit\'e de Savoie.

\begin{fmffile}{samplepics}
 
\end{fmffile}

\end{document}